\documentclass[
    onecolumn,
    preprintnumbers,
    amsmath,
    amssymb,
    prd,
    showpacs,
    showkeys
]{revtex4}
\usepackage{graphicx}%
\usepackage{amsmath}
\usepackage[normalem,normalbf]{ulem}
\usepackage{amssymb}
\usepackage{bm}
\usepackage{enumerate}

\begin{document}
\title{Zero mode in the time-dependent
  symmetry breaking of $\lambda\phi^4$ theory}
\author{Hyeong-Chan Kim}
\email{hckim@phya.yonsei.ac.kr}
\author{Jae Hyung Yee}%
\email{jhyee@phya.yonsei.ac.kr}
\affiliation{Dept. of Physics, Yonsei University, Seoul Republic
of Korea.
}%
\date{\today}%
\bigskip
\begin{abstract}
\bigskip

We apply the quartic exponential variational approximation to the symmetry
breaking phenomena of scalar field in three and four dimensions. We
calculate effective potential and effective action for the time-dependent
system by separating the zero mode from other non-zero modes of the scalar
field and treating the zero mode quantum mechanically. It is shown that
the quantum mechanical properties of the zero mode play a non-trivial role
in the symmetry breaking of the scalar $\lambda \phi^4$ theory.

\end{abstract}
\maketitle

\section{Introduction}
The variational approach for scalar field theory using the
Gaussian effective potential was well studied in
Refs.~\cite{stevenson,barns}, and references therein. The
renormalizability and initial value problems for the Gaussian
approximation are checked in ~\cite{cooper3,pi1}. Many authors
have studied the symmetry breaking phase structures in the large
$N$ approximation~\cite{baacke}. Nonequilibrium dynamics of
symmetry breaking~\cite{cooper2} and the second order phase
transition~\cite{spkim} have also been studied in the Gaussian
frame work.

On the other hand, it is known that the Gaussian approaches are
inappropriate to treat the time-dependent symmetry breaking of initially
symmetric states because of its own limits~\cite{cooper1}, that it cannot
be applied to a double well type potential for large dispersions. To
overcome this difficulty, many approaches have been attempted to go beyond
Gaussian methods by using quartic exponential ansatz~\cite{polley},
perturbative expansion around the Gaussian approximation~\cite{ghlee}, and
variation including higher excited states of Gaussian
approximation~\cite{cheetham}.

It is therefore important to understand the limit of the approaches based
on the Gaussian wave-functional and to develop consistent method going
beyond Gaussian approximation. Recently, the present authors have
developed a new quartic exponential type variational
approximation~\cite{kim1} which is suitable for systems with double well
type potentials in the quantum mechanical context. In this paper, by
applying the approximation to the zero mode of the scalar field, we found
the renormalized equations of motion which can describe the symmetry
breaking phenomena starting from the symmetric states of the scalar
$\phi^4$ theory in three and four dimensions.

The paper is organized as follows. First we separate the zero mode of the
scalar field from other modes in Sec. II and then calculate the formal
expression for the effective action. We then calculate the renormalized
effective potentials and actions for the (2+1) dimensional case in Sec.
III, and (3+1) dimensional case in Sec. IV. Finally, we summarize our
results and present some discussions in Sec. V.
\section{Mode separation of the self-interacting scalar field}
We first separate the zero-mode from other non-zero excitations
and present the concept of effective action adapted in the present
paper.

The Lagrangian of the scalar $\phi^4$ theory in $n+1$ dimensions is
\begin{eqnarray} \label{L}
L= \int d^n {\bf x} \left[ \frac{1}{2} \partial^\mu \phi(x^\nu)
\partial_\mu \phi(x^\nu) - \frac{1}{2}\mu^2(t) \phi^2(x^\nu)
 -\frac{\lambda}{4!} \phi^4(x^\nu)  \right] ,
\end{eqnarray}
where we explicitly included the volume integral so that we can
write the volume factor in zero mode part of the Lagrangian and
$\mu^2(t)$ asymptotically approaches to a negative value $\mu^2<0$
so that the system undergoes symmetry breaking. Let the theory be
defined in a box of volume $V$. Then, its zero mode can be
extracted by the following definitions:
\begin{eqnarray} \label{phi_0}
\phi(t) = \int d^n {\bf x} \phi({\bf x},0), ~~~ \psi({\bf x},t) =
\phi({\bf x},t) -
    \phi(t) .
\end{eqnarray}
The new field, $\psi$, satisfies $\int d^n{\bf x} \psi({\bf
x},t)=0$. The Lagrangian~(\ref{L}),
\begin{eqnarray} \label{L:1}
L&=& V\left[\frac{1}{2}\dot{\phi}^2(t)- \frac{\mu^2(t)}{2}
\phi^2(t)
   -\frac{\lambda}{4!} \phi^4(t) \right] \\
 && + \int d^n {\bf x} \left\{ \frac{1}{2}
  \partial^\mu \psi(x^\nu)
  \partial_\mu \psi(x^\nu) - \frac{1}{2} \left[\mu^2(t) +
  \frac{\lambda}{2} \phi^2(t)\right] \psi^2(x^\nu)
   -\frac{\lambda}{4!} \psi^4(x^\nu)- \frac{\lambda}{6}
   \phi(t) \psi^3(x^\nu) \right\} \nonumber ,
\end{eqnarray}
then, becomes a coupled Lagrangian of a quantum mechanical quartic
oscillator $\phi$ and a scalar field $\psi$ which has only
non-zero mode excitations.

The effective action for a time-dependent system is given by the
functional integral
\begin{eqnarray} \label{Gamma:1} \Gamma=
\int dt \langle \phi,\psi| i \partial_t - H |\phi,\psi\rangle ,
\end{eqnarray}
where the state $|\phi, \psi \rangle$ is the one that extremizes $\Gamma$.
The effective potential is given by the negative of its static limit.
Before we consider the quantum mechanical corrections, we give the tree
level analysis first. The effective mass of $\phi$ and $\psi$ are given by
\begin{eqnarray} \label{mass:0}
m^2_\phi (t) = \mu^2+ \frac{\lambda}{2V}\int d^n {\bf x}
\langle\psi^2({\bf x},t)\rangle , ~~m^2_\psi(t) = \mu^2 +
\frac{\lambda}{2} \langle \phi^2(t) \rangle ,
\end{eqnarray}
where $\langle A \rangle$ denotes the expectation value of $A$ for a
symmetric wave-functional $\Phi(\phi,\psi)$ given below. $\langle
\psi^2(x)\rangle$ usually become independent of $x^i$, the spatial
coordinates, due to the translational invariance of Green's function and
its spatial integral cancels the volume factor at the denominator in the
first equation of ~(\ref{mass:0}). In the presence of symmetry breaking,
we expect that the expectation value of the field $\langle
\phi^2(x,t)\rangle$ increases to a larger value $\displaystyle \phi^2 \sim
\frac{-\mu^2}{6\lambda}$. This allows the possibility of $\langle
\phi^2(t) \rangle$ having large enough value so that $m_\phi^2 <0$ and
$m_\psi^2>0$. This case is exactly what we are interested in in this
paper. Since the mass $m_\psi^2$ is positive definite, the
$\langle\psi^2\rangle$ cannot grow to larger value and confined to near
zero value. The stable equilibrium of $\phi$ is at $\pm
\sqrt{\frac{6}{\lambda} [-\mu^2-\frac{\lambda}{2V}\int d^n{\bf x} \langle
\psi^2 \rangle] }$. Therefore, $\phi$ stays in the degenerate ground state
since the large volume factor $V$ makes the potential wall between the
positive and negative minima infinitely high.

These discussions justify the use of Gaussian approximation for $\psi$
field. On the other hand, we use quartic exponential approximation for the
zero mode, $\phi$, to include better quantum mechanical effects. To
describe the evolution of the symmetric state we use the trial wave
functional of the form
\begin{eqnarray} \label{wf:1}
\Phi[\phi,\psi]&=& N \exp\left\{ -
\frac{1}{2}\left[\frac{1}{2g^2(t)} + i \pi(t) \right] \phi^4
+\left[\frac{x(t)}{g(t)} +i p(t) \right] \phi^2  \right. \\
&-& \left. \int_{\bf x y}\psi({\bf x})\left[\frac{G^{-1}({\bf
x,y};t)}{4}- i \Pi({\bf x,y};t) \right]\psi({\bf y}) \right\} ,
\nonumber
\end{eqnarray}
where $\int_{\bf x}=\int d^n{\bf x}$ and we use the unit which
makes $\hbar=1$. The use of this symmetric trial wave-functional
simplifies the computation since the contribution from the
interaction term, $\phi \psi^3$, in the Lagrangian~(\ref{L:1})
vanishes. Following Ref.~\cite{kim1} we define some physical
quantities of the zero mode:
\begin{eqnarray} \label{qy}
q^2(t) = \langle \phi^2(t) \rangle, ~~ y(t)= \frac{\langle
\phi^4(t) \rangle }{(\langle \phi^2(t) \rangle)^2 }.
\end{eqnarray}
By introducing the integral~\cite{kim1}
\begin{eqnarray} \label{integ}
f(x)= \frac{1}{\sqrt{g}}\int_{-\infty}^{\infty}dQ \exp
\left(-\frac{Q^4}{2g^2}+ \frac{2xQ^2}{g} \right)
=\frac{\pi}{\sqrt{2}} |x|^{1/2}e^{x^2}\left[I_{-1/4}(x^2)+ sgn(x)
I_{1/4}(x^2)\right],
\end{eqnarray}
we get
\begin{eqnarray} \label{qy}
q^2(t)=\frac{g f'}{2 f}, ~~~ y(t)=\frac{1+ 2 x f'/f}{f'^2/(2f^2)}.
\end{eqnarray}
We additionally introduce the notation $\displaystyle
\frac{d\eta}{dy}\equiv D=\frac{1}{4} \sqrt{\frac{1+Y}{y(3-y)}}$,
$\displaystyle Y= \frac{2x f'(x)}{f(x)}$ as in Ref.~\cite{kim1}.

Using these, we write the effective action~(\ref{Gamma:1}) as a functional
of $q(t)$, $p(t)$, $y(t)$, $\pi(t)$, $\Pi({\bf x,y};t)$, and $G({\bf
x,y};t)$:
\begin{eqnarray} \label{Gamma:2}
\Gamma &=&  \int dt\left\{ \frac{yq^4\dot{\pi}}{2}- q^2 \dot{p}-
  2y\left[y- \frac{y-3}{Y+1}\right]\frac{q^6 \pi^2}{V}
  -2 \frac{q^2 p^2}{V} + yq^4
  \frac{\pi p}{V} -V V_{eff}(q,y) \right. \\
&+& \left. \int_{\bf xy}\Pi \dot{G}-2 \int_{\bf xyz}\Pi G \Pi-
  \int_{\bf x}\left[\frac{1}{8}G^{-1}({\bf x,x})
   - \frac{1}{2}\nabla^2_{\bf x}
  G({\bf x,y};t)|_{\bf x=y}+ \frac{1}{2}V^{(2)}
   G({\bf x,x};t) \right] \right.  \nonumber \\
   & -&\left.
  \frac{\lambda}{8}\int_{\bf x}
   G({\bf x,x};t)G({\bf x,x};t) \right\} , \nonumber
\end{eqnarray}
where $\displaystyle V^{(2)}= \mu^2+ \frac{\lambda}{2}q^2(t)$ and
$V_{eff}(q,y)=\displaystyle \frac{V_F(y)}{8 V^2 q^2}+
\frac{\mu^2}{2} q^2 + \frac{\lambda}{4!} y q^4$. We do not need
the explicit form of $V_F$, instead, it is enough to know that
$V_F$ becomes divergent at $y=1$. Therefore, in large $V$ limit,
\begin{eqnarray} \label{Veff}
V_{eff} = \left\{ \begin{tabular}{ll}
  $\displaystyle \frac{1}{2}\mu^2 q^2 + \frac{\lambda}{4!}
     y q^4$ , &  ~~$y\geq 1$ and $q \geq 0$ ,\\
  $\infty$ ,   & ~~otherwise.
\end{tabular}\right.
\end{eqnarray}
Note that $V_{eff}$ is finite even in the limits of zero
dispersion $q^2 \rightarrow 0$ or $y \rightarrow 1$.

After solving the $p$ and $\pi$ equations, and using the
translational invariance of $G$ equation of motion we get
\begin{eqnarray} \label{Gamma:3}
\Gamma[q,y;G,\Pi] &=& \int dt d^n{\bf
x}\left\{\left[\frac{q^2\dot{\eta}^2}{2}
     + \frac{\dot{q}^2}{2} - V_{eff}(q,y)\right] +
  \int_{\bf k}\left[\Pi({\bf k},t) \dot{G}({\bf k},t)
      -2  \Pi^2({\bf k},t) G({\bf k},t)
   \right. \right. \\
&-& \left. \left.
  \frac{1}{8}G^{-1}({\bf k},t) - \frac{1}{2}
   ({\bf k}^2
 + V^{(2)}) G({\bf k},t) -
  \frac{\lambda}{8}G({\bf k},t)
   \int_{\bf q} G({\bf q},t) \right] \right\} , \nonumber
\end{eqnarray}
where $G({\bf k},t)$ and $\Pi({\bf k},t)$ are the Fourier
transforms of $G({\bf x,y};t)$ and $\Pi({\bf x,y};t)$:
\begin{eqnarray} \label{fourier}
G({\bf x,y};t)= \int_{\bf k} G({\bf k},t) e^{i {\bf k}\cdot ({\bf
x-y})},~~ \Pi({\bf x,y};t)= \int_{\bf k} \Pi({\bf k},t) e^{i {\bf
k}\cdot ({\bf x-y})},
\end{eqnarray}
where $\displaystyle \int_{\bf k}= \int \frac{d^n {\bf
k}}{(2\pi)^n}$.

The time-dependent variational equations are given by
\begin{eqnarray} \label{eqs}
&& \frac{d}{dt}(q^2 D\dot{y})+ \frac{\lambda}{4! D}q^4=0 , \\
   \label{ddq:uR}
&&\ddot{q}+[m^2(t)-\dot{\eta}^2(t)]q+ \frac{\lambda}{6}(y-3)q^3=0, \\
&& \ddot{G}({\bf k},t)= \frac{1}{2}G^{-1}({\bf k},t)+ \frac{1}{2}
G^{-1}({\bf k},t)\dot{G}^2({\bf k},t)-2
   m^2(t) G({\bf k},t) , \label{gap:1}
\end{eqnarray}
where
 \begin{eqnarray} \label{m:t}
 m^2(t) &=& \mu^2(t)+
\frac{\lambda}{2} q^2(t)+ \frac{\lambda}{2} \int_{\bf k} G({\bf
k},t) .
\end{eqnarray}
The potential divergences come from the integral $\int_{\bf
k}G({\bf k},t)$. $D(y)$ becomes divergent at $y=1$ such that
$\displaystyle D(y) \sim \frac{1}{4}\frac{1}{\sqrt{y-1}}$ as $y
\rightarrow 1$.

\section{Effective action in (2+1)-dimensions}
Since the Gaussian effective potential for the scalar $\phi^4$ field
theory in (2+1) dimensions shows symmetry breaking, (2+1) dimensional
theory is the best place to test our approximation method that uses
zero-mode separation and quartic Gaussian trial wave-functional. During
these calculations we assume $m_\phi^2\geq 0$ always. Therefore, the
non-zero modes stay always in symmetry restored states with zero
expectation value, $\langle \psi({\bf x},t)\rangle=0$, and the zero mode
is in a symmetry broken state.

\subsection{Effective potential}
In this subsection, we shall briefly discuss how the static effective
potential is renormalized in the quartic exponential treatment of the zero
mode.

The definition of effective potential in this paper is slightly different
from the standard one. In this paper, we separate the field into the zero
mode and non-zero mode parts, then, we integrate out the non-zero mode (by
solving the gap equation~(\ref{gap:1}) of the non-zero modes) to get the
effective potential for the zero mode. This definition gives slightly
different effective potential, since the zero mode is treated quantum
mechanically in this paper, while the traditional method treat it as a
background classical field. In spite of these conceptual differences the
calculational procedure is similar to that of the Gaussian approximation.

The effective potential per unit volume in the present
approximation is given by
\begin{eqnarray} \label{V:1}
V_{eff}(y,q;G(q))= V_{eff}(y,q)+\frac{1}{4}\int_{\bf k}G^{-1}({\bf
k})- \frac{\lambda}{8} \left[ \int_{\bf k}G({\bf k})\right]^2.
\end{eqnarray}
The potential is minimized at $y=1$ and $G$ is the solution of the
gap equation:
\begin{eqnarray} \label{gap}
\frac{1}{4}G^{-2}({\bf k})={\bf k}^2+ m^2={\bf k}^2+ \mu^2+
\frac{\lambda}{2}q^2+ \frac{\lambda}{2}\int_{\bf k} G({\bf k}) ,
\end{eqnarray}
which gives $G(q)$. The possible source of divergence comes from
the integral
\begin{eqnarray} \label{G}
\int_k G({\bf k})
   = I_0(m)=\int_{\bf k} \frac{1}{2 \sqrt{{\bf k}^2+ m^2 }} .
\end{eqnarray}
The present form of $G$ is the same as that of the Gaussian
approximation of Ref.~\cite{stevenson} with a slight modification
that $q^2$ is now the expectation value of dispersion of
zero-mode. This divergence in $I_0(m)$ is absorbed in $\mu^2$ by
defining the renormalized mass
\begin{eqnarray} \label{mR:3}
m_R^2= \mu^2+\frac{\lambda}{2}\left.\int_{\bf k} G({\bf
k})\right|_{q=0},
\end{eqnarray}
with the help of the formula
\begin{eqnarray} \label{I0}
I_0(m)- I_0(m_R)= -\frac{m-m_R}{4\pi} .
\end{eqnarray}
Then, the gap equation~(\ref{gap}) can be written to give the
ratio of the mass with respect to the renormalized mass for
general $q$
\begin{eqnarray} \label{x:q}
\sqrt{x}\equiv \frac{m}{m_R}= -\frac{\lambda}{16\pi m_R}+
\left[\left(\frac{\lambda}{16\pi m_R}+1 \right)^2+ \frac{\lambda
}{2 m_R^2}q^2 \right]^{1/2} .
\end{eqnarray}

The effective potential in (2+1) dimensions can then be calculated to be
\begin{eqnarray} \label{V:3d}
V(q,y)&=&\frac{1}{2} m_R^2 q^2+ \frac{\lambda}{4!}yq^4-
\frac{m_R^3}{24 \pi} (\sqrt{x}-1)^2\left[2 \sqrt{x}+ 1 +
\frac{3\lambda}{16\pi m_R}\right], ~~q\geq  0 , \label{x:q}
\end{eqnarray}
which exactly reproduces the result of Ref.~\cite{stevenson} if
$y=1$.
Note also that the potential for $\lambda \rightarrow \infty$
becomes
\begin{eqnarray} \label{V:infty}
\frac{V(q,y)}{m_R^3} \simeq -\frac{\lambda}{24 m_R^3}(3-y) q^4 +
\frac{q^2}{2m_R}+ \frac{2 \pi q^4}{m_R^2}+ \frac{8\pi q^6}{3m_R^3}
+O(1/\lambda).
\end{eqnarray}
This form shows two things: First is that there is symmetry
breaking for $y<3$ for large enough $\lambda/m_R$ always.
Therefore we have a critical value $\lambda/m_R=\hat{\lambda}_c$
such that if $\lambda/m_R > \hat{\lambda_c}$ there is symmetry
breaking for $y<3$. Second is that the potential for $q$ at $y=3$
does not have symmetry breaking form for any large value of
$\lambda/m_R$. This fact shows that there is a critical value
$y=y_c(\lambda)$, such that for $y<y_c<3$ and $\lambda/m_R>
\hat{\lambda_c}$ symmetry breaking occurs. We draw the effective
potential for various $y$ values as a function of $q$ in Fig. 1.
\begin{figure}[htbp]
\begin{center}
\includegraphics[width=.6\linewidth,origin=tl]{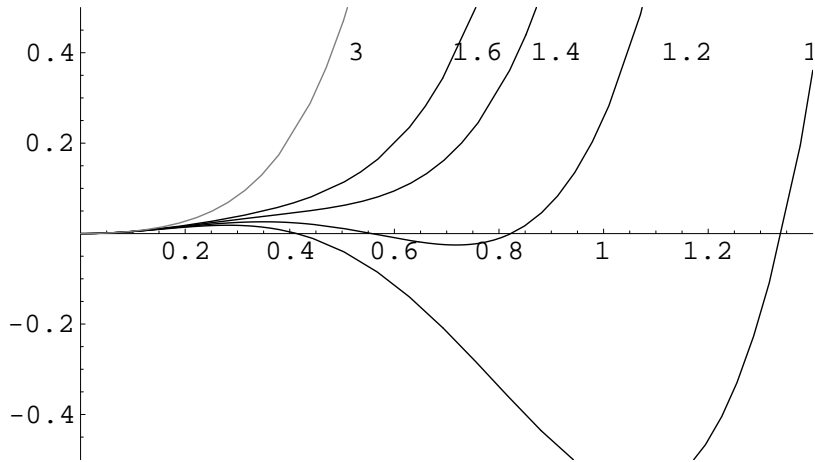}\hfill%
\end{center}
\caption{The effective potential $V/m_R^3$ as a function of
$\bar{q}=q/\sqrt{m_R}$. In this figure we set $\lambda=4 m_R$,
$q_R=0$, and the curves represents the cases of $y=1, 1.2, 1.4,
1.6, 3$.} \label{qx:fig}
\end{figure}
From Fig. 1 we can deduce the dynamics of a zero-mode wave-packet with the
initial values $(q_0,y_0=3)$. The dynamics of the system is described by
two processes. First is the dynamics until $y$ decreases to $y=y_c$.
During this process, the ground state for $q$ is at $q=0$. The dispersion
$q^2$ will oscillates around zero. The second process begins when $y$
decreases to a value below $y_c$, so that the potential has symmetry
breaking form. From this time, $q$ increases until it takes stable
equilibrium value and $y$ will oscillates around $y=1$. During these
processes, the zero mode continually interacts with the non-zero modes and
may loose much of its energy. If one wants to understand these dynamical
processes better, we need to calculate the effective action.

\subsection{Renormalization of the effective action}
It is known in Ref.~\cite{pi1} that not all states are allowed as the
initial states in the Gaussian variational approximation. What we consider
in this paper is that the evolution of symmetric initial states. Even
though we treat the zero mode separately, our method uses Gaussian
approximation for the non-zero modes. This implies that the
renormalizability condition for the time-dependent variational equations
in Ref.~\cite{pi1} applies also to the present case. Thus we simply
outline the initial condition and then write down the equations of motion.
We choose the initial state of non-zero modes as a Gaussian with
$G(x,y;0)=G(x,y)$, and $\Pi(x,y;0)= \Pi(x,y)$, and the initial state of
the zero mode as a quartic exponential with $q=q_0$, $y=y_0$, $p(0)=0$,
$\pi(0)=0$ in Eq.~(\ref{wf:1}). Here we assume that there are no
$\psi(x,t)-\phi(t)$ correlation in the initial state for simplicity. Then
the non-zero overlapping condition for the initial states with the vacuum,
prescribed in Ref.~\cite{pi1}, are given by
\begin{eqnarray} \label{G3:large}
\lim_{k\rightarrow \infty} G(k,0)= \frac{1}{2k} \left(1+ \frac{g
\cos \alpha(k)}{k} +\frac{\bar{m}^2}{k^2} \right),~~ \lim_{k
\rightarrow \infty} \dot{G}({\bf k},0)= \frac{B\cos \beta(k)}{k} +
\frac{A}{k^2},
\end{eqnarray}
where  $\alpha$ and $\beta$ are non-oscillatory and $g$, $A$, $B$,
and $\bar{m}$ are $k$-independent constants.

The renormalization condition~(\ref{mR:3}) can be generalized to
the time dependent case:
\begin{eqnarray} \label{mR:t}
m^2_R(t)= \mu^2(t)+ \frac{\lambda}{2} \left.\int_{\bf k} G_V({\bf
k},t) \right|_{q=0} ,
\end{eqnarray}
where $\displaystyle G_V({\bf k},t)=\frac{1}{2[{\bf k}^2+
\mu^2(t)+ \frac{\lambda}{2}q^2 +\frac{\lambda}{2}\int_{\bf k}G_V
]^{1/2}}$ is the value of $G$ in the instantaneous vacuum for a
given $q$. The finite time-dependent mass is
\begin{eqnarray} \label{m:t}
m^2(t) &=& \left[\mu^2(t)+ \frac{\lambda}{2}q^2(t) +
    \frac{\lambda}{2}\int_{\bf k}
    G_V({\bf k},t)\right] + \frac{\lambda}{2} \int_{\bf k}
 \left[G({\bf k},t)-G_V({\bf k},t)\right] \\
 &=& m_V^2(t)+ \tilde{m}^2(t) , \nonumber
\end{eqnarray}
where
\begin{eqnarray} \label{mmm}
m_V^2(t) &\equiv& \mu^2(t)+ \frac{\lambda}{2}q^2(t) +
\frac{\lambda}{2}\int_{\bf k} \frac{1}{2({\bf k}^2 + m_V^2(t))^{1/2}} \\
&=& m_R^2(t)+ \frac{\lambda}{2}q^2(t) +
\frac{\lambda}{2}[I_0(m_V)-I_0(m_R)] . \nonumber
\end{eqnarray}
Using Eq.~(\ref{I0}) we get
\begin{eqnarray} \label{m_V}
m_V(t)=-\frac{\lambda}{16\pi}+
\left[\left(\frac{\lambda}{16\pi}\right)^2+ m_R^2(t)+
\frac{\lambda}{8\pi} m_R(t) + \frac{ \lambda}{2} q^2(t)
\right]^{1/2} .
\end{eqnarray}
One can show that the difference of mass-squared
$\tilde{m}^2=m^2-m_V^2$,
\begin{eqnarray} \label{tildem}
\tilde{m}^2(t)&\equiv & \frac{\lambda}{2} \int_{\bf k} [G({\bf
k},t)-G_V({\bf k},t)] ,
\end{eqnarray}
is finite by using the explicit asymptotic form of $G$ in
Eq.~(\ref{G3:large}) at $t=0$. Now we need to show that $\tilde{m}^2(t)$
is finite for all $t$. This can be done by solving the renormalized
equation of motion for $G$. We do not write this calculation explicitly
here since it is a mere repetition of Eq. (4.11) of Ref.~\cite{pi1}.

The renormalized equations of motion for $q$ and $G({\bf k},t)$
become
\begin{eqnarray}
&&\ddot{q}+ \left[m_V^2(t)+\tilde{m}^2(t) \right]q +
\frac{\lambda}{6}yq^3= \dot{\eta}^2 q, \label{ddq:R3} \\
&&\ddot{G}({\bf k},t)=\frac{1}{2}G^{-1}({\bf k},t)+
\frac{1}{2}G^{-1}({\bf k},t)\dot{G}^2({\bf k},t)-2\left[{\bf k}^2+
m_V^2(t)+\tilde{m}^2(t) \right]G({\bf k},t) . \label{ddG:R}
\end{eqnarray}
Note that near $y=1$, $D \sim 1/(4\sqrt{y-1})$ diverges. Therefore
$q^2 \dot{\eta}$ becomes almost constant in this region. Then, the
driving term of $q$ in the right hand side of ~(\ref{ddq:R3})
becomes divergent as $1/q^3$ for small $q$. This may enable $q$ to
take over the low potential wall located between $q=0$ and the
true vacuum in the effective potential in Fig. 1.

We have thus shown that for the initial states, which belong to the Fock
space built on the vacuum, the time-dependent variational equation is made
finite by the static renormalization used in the vacuum sector in (2+1)
dimensions. The present method naturally incorporates the quantum
mechanical correction by $y$ to the classical potential which is given by
$y=1$. A notable conceptual difference coming from the quantum treatment
of the zero mode of the (2+1) dimensional scalar field theory is that a
Gaussian wave-packet is not a minimum energy packet even for unbroken
non-degenerated vacuum. From Eq.~(\ref{eqs}), we see that $y$ is stable
only at $y=1$, the two delta-like packets limit. In Eq.~(\ref{ddq:R3}) $y$
plays dual role as a potential which roll down $q$ for $y>y_c$, then, as
an external force which boosts $q$ to a larger value for $y<y_c$. This
quantum correction from $y$ contributes to $G$ only indirectly through
$q$.

\section{Effective action in (3+1) dimensions}
In (3+1) dimensions, the situation is quite different from the (2+1)
dimensional case. This is because of the renormalization condition, which
demands $\lambda \rightarrow 0_-$ which in turn leads to the conclusion
that there is no phase transition in the Gaussian effective potential of
four dimensional scalar $\phi^4$ theory. At the present case, if $\lambda
\rightarrow 0_-$, the effective potential looses $y$ dependent term and
$y$ dynamics decouples from the rest of the effective potential. If we
starts from the effective action, however, $y$ dynamics leads to a
repulsive potential proportional to $1/q^{2}$, which may give rise to
symmetry breaking form of the potential even though some interpretational
difficulties remain.

\subsection{effective potential in the standard method}
Let us consider the effective potential first. Since the present
form for $G$ in Eq.~(\ref{gap}) is the same as that of the
Gaussian approximation with a slight modification that $q^2$ is
now the expectation value of dispersion, the renormalized
quantities are similarly defined as in Ref.~\cite{pi1}:
\begin{eqnarray} \label{ren}
\frac{\mu_R^2}{\lambda_R}&=& \frac{\mu^2}{\lambda}+
\frac{1}{2}I_1, ~~ I_1 \equiv \int_k \frac{1}{2k},
~~ \\
\frac{1}{\lambda_R}&=& \frac{1}{\lambda}+ \frac{1}{2} I_2(M), ~
I_2(M)\equiv \frac{1}{M^2}\int_k \left[\frac{1}{2k}- \frac{1}{2
(k^2+ M^2)^{1/2}}\right] ,
\end{eqnarray}
where $M$ is an arbitrary mass scale, at which the renormalization
is performed. The gap equation~(\ref{gap}) becomes a kind of mass
renormalization formula,
\begin{eqnarray} \label{m}
m^2= \mu_R^2+ \frac{\lambda_R}{2}q^2 + \frac{\lambda_R}{2}
m^2[I_2(M)- I_2(m)] = \mu_R^2+\frac{\lambda_R}{2}q^2
+\frac{\lambda_R}{32\pi^2}m^2 \ln \frac{m^2}{M^2} .
\end{eqnarray}
We obtain a finite expression of the effective potential:
\begin{eqnarray} \label{Veff:4}
V(y,q)= \frac{m^4-\mu_R^4}{2 \lambda_R} + \frac{m^4}{64 \pi^2}
    \left(\ln \frac{M^2}{m^2}- \frac{1}{2} \right) +
    \frac{y-2}{24}\lambda q^4 ,
\end{eqnarray}
where we have adjusted a constant $-\mu^4/2\lambda$ in the limit of
infinite cutoff, $\lambda \rightarrow 0_-$. At $y=1$, this effective
potential~(\ref{Veff:4}) and the gap equation~(\ref{m}) reproduces the
results of Pi and Samiullah~\cite{pi1} and Stevenson~\cite{stevenson},
where it was shown that the scalar $\phi^4$ model does not have symmetry
breaking in $(3+1)$ dimensions in the Gaussian approximation. As the
coupling $\lambda$ goes to zero, the $y$ dependence in the quartic term
disappears. We interpret this potential as the effective potential for
$q$. The explicit form of the effective potential with respect to $q$ can
be obtained by replacing $m^2$ in (\ref{Veff:4}) by using Eq.~(\ref{m}).
\subsection{Renormalization of the effective action in 3+1 dimensions}
The condition for the initial states~\cite{pi1}, in (3+1)
dimensions, are given by
\begin{eqnarray} \label{G:large}
\lim_{k\rightarrow \infty} G(k,0)= \frac{1}{2k} \left(1-
\frac{\bar{m}^2}{2k^2}+ \frac{g \cos \alpha(k)}{k^2} \right), ~~~
\lim_{k\rightarrow \infty}\dot{G}(k,0)= \frac{A+ B \cos
\beta(k)}{k^2} ,
\end{eqnarray}
with nonoscillatory $\alpha$ and $\beta$ and $k$-independent
constants $g$, $A$, $B$, and $\bar{m}$, the last being a mass
parameter that we shall specify shortly.

We generalize the renormalization condition to the time-dependent
$\mu^2$ as follows:
\begin{eqnarray} \label{ren:t}
&& \frac{\mu^2_R(t)}{\lambda_R}= \frac{\mu^2(t)}{\lambda}+
     \frac{1}{2} I_1, \\
&&\frac{1}{\lambda_R}= \frac{1}{\lambda} + \frac{1}{2} I_2(M).
\end{eqnarray}
This renormalization condition implies $\lambda \rightarrow 0_-$.
Therefore, the first equation in ~(\ref{eqs}) for $y$ gives $q^2
\dot{\eta}=C$, a constant of motion. With this condition, we remove $y$
dependence and write the second equation of (\ref{eqs}) as
\begin{eqnarray} \label{ddq:2}
&&\ddot{q}+m^2(t)q -\frac{C^2}{q^3}=0.
\end{eqnarray}
Then, the effective action~(\ref{Gamma:3}) we are to solve becomes
\begin{eqnarray} \label{ac:3}
\Gamma'[q,G,\Pi] &=& \int dt d^3x\left\{
\left[\frac{\dot{q}^2}{2}-\frac{C^2}{2q^2}
     - \frac{1}{2} \mu^2 q^2 \right] +
  \int_{\bf k}\left[\Pi({\bf k},t) \dot{G}({\bf k},t)
      -2  \Pi^2({\bf k},t) G({\bf k},t)
   \right. \right. \\
&-& \left. \left.
  \frac{1}{8}G^{-1}({\bf k},t) - \frac{1}{2}
   ({\bf k}^2
 + V^{(2)}) G({\bf k},t) -
  \frac{\lambda}{8}G({\bf k},t)
   \int_{\bf q} G({\bf q},t) \right] \right\} . \nonumber
\end{eqnarray}

In the present approach, $q^2(t)=\langle \phi^2(t)\rangle$ cannot
vanish if $C \neq 0$, and should be dynamical. This fact gives
slight difference to the renormalization procedure from that of
Ref.~\cite{pi1}. The finite time-dependent mass becomes
\begin{eqnarray} \label{m:t}
m^2(t)
 &=& \left[\mu^2(t)+ \frac{\lambda}{2}q^2(t) +
    \frac{\lambda}{2}
    I_0(m_V(t))\right] + \frac{\lambda}{2} \int_k
 \left[G(k,t)-G_V(k,t)\right] \\
 &=& m_V^2(t)+ \tilde{m}^2(t) , \nonumber
\end{eqnarray}
therefore
\begin{eqnarray} \label{mmm}
m_V^2(t) &\equiv& \mu^2(t)+ \frac{\lambda}{2}q^2(t) +
\frac{\lambda}{2}\int_k \frac{1}{2(k^2 + m_V^2(t))^{1/2}}=
\mu_R^2(t)+ \frac{\lambda_R}{2}q^2(t) +
\frac{\lambda_R}{32\pi^2}m_V^2(t) \ln \frac{m_V^2(t)}{M^2},
\end{eqnarray}
where
\begin{eqnarray} \label{tildem}
\tilde{m}^2(t)&\equiv & \frac{\lambda}{2} \int_k
[G(k,t)-G_V(k,t)]  \\
&=& \frac{\lambda_R}{2}\int_k [G(k,t)-G_V(k,t)] +
\frac{\lambda_R}{2} \tilde{m}^2(t) I_2(M) , \nonumber
\end{eqnarray}
with $\mu_R(t)$ becoming independent of time after the symmetry
breaking process is over. After rearranging these terms we get
\begin{eqnarray} \label{m:ren}
m^2(t)&=& \mu_R^2(t)+ \frac{\lambda_R}{2}u^2(t) +
\frac{\lambda_R}{2} q^2(t), \\
u^2(t)& =& \frac{1}{16\pi^2}m_V^2(t) \ln \frac{m_V^2(t)}{M^2}+
\int_k[G(k,t)-G_V(k,t)]+ \tilde{m}^2(t) I_2(M) . \nonumber
\end{eqnarray}
The finiteness of $u^2(t)$ at $t=0$ gives $\bar{m}^2= m^2(0)=
m_V^2(0) + \tilde{m}^2(0)$, where we use the asymptotic
form~(\ref{G:large}) for $G$ at $t=0$. In the time-independent
limit, $u^2= \frac{1 }{16 \pi^2} m^2 \ln \frac{m^2}{M^2}$. The
renormalized set of equations of motion is given by
\begin{eqnarray}
&&\ddot{q}+ \left[\mu_R^2(t)+\frac{\lambda_R}{2}u^2(t) \right]q +
\frac{\lambda_R}{2}q^3- \frac{C^2}{q^3}=0, \label{ddq:R} \\
&&\ddot{G}=\frac{1}{2}G^{-1}+
\frac{1}{2}G^{-1}\dot{G}^2-2\left[k^2+ \mu_R^2(t)+
\frac{\lambda_R}{2}u^2(t) + \frac{\lambda_R}{2} q^2(t)\right]G .
\label{ddG:R}
\end{eqnarray}
From the equation of motion~(\ref{ddq:R}) one finds that the minimum of
the potential is given  at the point $q_v$ determined by
\begin{eqnarray} \label{q:min}
0=\left. \frac{\partial V'}{\partial
q}\right|_{q=q_v}=q_v\left[\mu_R^2+\frac{\lambda_R}{2} u^2 +
\frac{\lambda_R}{2}q_v^2 - \frac{C^2}{q_v^4}\right],
\end{eqnarray}
which signals that $q_v=0$ cannot be a minimum of the potential if
$C \neq 0$.

Finally, let us examine the effective potential again after integrating
out the equation of motion for $y$ with non-zero $C$. Let us assume that
the system is quasi-static so that $G$ is described by its vacuum value.
Since the potential is divergent at $q=0$, we define the renormalized mass
at $q=q_v$ where the potential takes minimum:
\begin{eqnarray} \label{mR:4}
m_R^2=\mu_R^2+ \frac{\lambda_R}{2}q^2_v+ \frac{\lambda_R}{32\pi^2}
m_R^2 \ln \frac{m_R^2}{M^2}.
\end{eqnarray}
By setting the arbitrary mass scale $M=m_R$, we get
$m^2_R=\mu_R^2+ \frac{\lambda_R}{2} q_v^2$. Following the
calculation of Stevenson, the mass-square ratio $x=m^2/m_R^2$ is
given by
\begin{eqnarray} \label{m:q}
\kappa \left(x-1-\frac{\bar{q}}{\kappa}  \right) =
 x\ln x,
\end{eqnarray}
where $\displaystyle \kappa= \frac{32 \pi^2}{\lambda_R}$ and
$\displaystyle \bar{q}= 16\pi^2\frac{q^2-q_v^2}{m_R^2}$. The effective
potential is, then, given by
\begin{eqnarray} \label{V:eff2}
V'(x,\Phi)=\frac{64 \pi^2 V_{eff}}{m_R^4}(x,q)= \frac{c^2}{\Phi^2} +\kappa
(x^2-1)- x^2 \left(\ln x + \frac{1}{2} \right) ,
\end{eqnarray}
where $\Phi^2= 16 \pi^2 q^2/m_R^2$ and $c=32\pi^2 C/m_R^3$. Unlike the
previous result~(\ref{Veff:4}), this potential has the symmetry breaking
form, due to $C^2/\Phi^2$ term, which comes from $y$ dynamics.
\begin{figure}[htbp]
\begin{center}
\includegraphics[width=.6\linewidth,origin=tl]{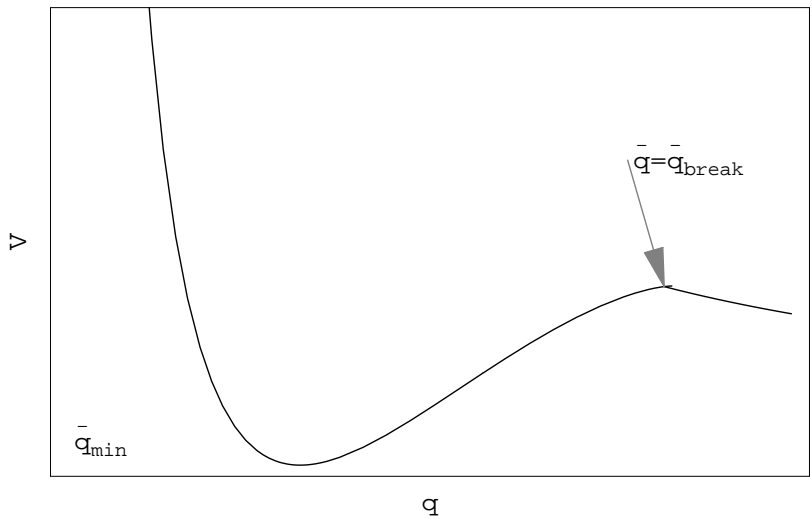}\hfill%
\end{center}
\caption{The effective potential as a function of $\bar{q}$.
Details of the analysis for $\bar{q}_{break}$ are the same
as~\cite{stevenson} and $\bar{q}_{min}=-\Phi_v^2$. At this point
$q^2=0$.} \label{vq:fig}
\end{figure}

The presence of a symmetry breaking potential due to $y$ dynamics appear
to be surprising. However, many authors have been searching for a second
order phase transition in the scalar $\phi^4$ theory in (3+1) dimensions
through the study of non-equilibrium dynamics~\cite{spkim,cormier}. In
these studies the spinodal instability leads to better understanding of
the increase in the correlation length leading to the possibility of
symmetry breaking. In this sense, the presence of symmetry breaking
potential is not surprising, but more careful study is need.
The reason is that our wave-functional ansatz~(\ref{wf:1}) cannot describe
regions with $y>3$. Therefore, we cannot predict anything for $y>3$. In
this sense, more careful study with the trial wave-functional, that
include the region with $y>3$, is needed to clarify the issue of the
existence of symmetry breaking in the scalar $\phi^4$ theory in (3+1)
dimensions. One of methods to consider $y>3$ is to include the excited
states of~(\ref{wf:1}) as suggested in Ref.~\cite{kim1}.

\section{Summary and discussions}
We have calculated the effective action of a self-interacting scalar field
in three and four dimensions with the use of quartic exponential
wave-functional ansatz for zero mode.

In (2+1) dimensions, we have calculated the renormalized effective
potential and action. It is shown that the symmetry breaking occurs for
$\lambda/m_R >\hat{\lambda}_c$ as in the Gaussian case. The effective
potential is dependent on the shape, $y$, of wave-function of the zero
mode, so that there is a critical value, $y_c$, such that there is no
symmetry breaking if $y>y_c$. Especially, if $y=3$ there is no symmetry
breaking for any value of $\lambda/m_R$. The shape of the symmetry
breaking potential has double well type of the 1st order transition. If
the renormalized mass is defined at finite $q_R$,
\begin{eqnarray} \label{mR}
m_R^2= \mu^2+ \frac{\lambda}{2}q_R^2
+\frac{\lambda}{2}\left.\int_{\bf k} G({\bf k})\right|_{q=q_R},
\end{eqnarray}
then, the ratio of the mass for general $q$ and the renormalized
mass is given by
\begin{eqnarray} \label{x:q}
\sqrt{x}\equiv \frac{m}{m_R}= -\frac{\lambda}{16\pi m_R}+
\left[\left(\frac{\lambda}{16\pi m_R}+1 \right)^2+ \frac{\lambda
}{2 m_R^2}(q^2-q_R^2) \right]^{1/2}.
\end{eqnarray}
This definition allows the possibility that $m^2|_{q=0} \leq m_R^2$, and
$m^2|_{q=q_{min}}=0$ at
\begin{eqnarray} \label{m0:q}
\frac{q^2_{\rm min}}{m_R}= \frac{q_R^2}{m_R}-\frac{2 m_R}{\lambda}
- \frac{1}{4\pi}.
\end{eqnarray}
If we set $q_{\rm min}=0$, we get second order transition and
$\frac{\lambda}{2}q_R^2=  m_R^2 + \lambda m_R/(8\pi)$. The
effective potential in (2+1) dimension in this case becomes
\begin{eqnarray} \label{V:3d}
V(q,y)&=&-\frac{\lambda m_R}{16 \pi}q^2+ \frac{\lambda}{4!}yq^4-
\frac{m_R^3}{24 \pi} (\sqrt{x}-1)^2\left[2 \sqrt{x}+ 1 +
\frac{3\lambda}{16\pi m_R}\right]  . \label{x:q}
\end{eqnarray}
This is a new possibility for the (2+1) dimensional $\phi^4$ scalar field
theory since the center $q=0$ of this potential is an unstable maxima now.

We also have calculated the renormalized effective action
$\Gamma[q(t),y(t);G,\Pi]$ and then the effective potential $V(q,y)$ as its
static limit in (3+1) dimensions. The $y-$dependent term in $V(q,y)$
disappears after renormalization due to the renormalization condition
$\lambda \rightarrow 0_-$, and it leads to the result that the symmetry
breaking does not occur in 4-dimensional $\phi^4$ theory as in the case of
the Gaussian approximation.

Since $y-$dependence decouples from the rest during the renormalization
process, one can integrate out the $y-$dynamics from the effective action
$\Gamma[q(t),y(t);G,\Pi]$ to obtain a new effective action
$\Gamma'[q(t);G,\Pi]$ before one takes the static limit. The new effective
potential $V'(q)$ obtained by the static limit of $\Gamma'[q(t);G,\Pi]$
has the symmetry breaking form. To take this result seriously, however, we
need more careful study since in our trial wave-functional~(\ref{wf:1})
$y$ cannot take values larger than $3$.

\begin{acknowledgments}
This work was supported in part by Korea Research Foundation under
Project number KRF-2001-005-D2003 (H.-C.K. and J.H.Y.).
\end{acknowledgments}
\vspace{3cm}

\end{document}